\newif\ifhide

\documentclass{ws-fnl}

\usepackage{xcolor}
\usepackage[verbose]{hyperref}
\hypersetup{colorlinks=false,allbordercolors=blue,pdfborderstyle={/S/U/W 1}}

\usepackage{amsmath,amsfonts,amssymb}
\usepackage{graphics}
\usepackage{braket}

\renewcommand{\d}{\mathrm{d}}

\begin{document}

\markboth{Brian R. La Cour}{Wigner function negativity in a classical model of quantum light}


\title{Wigner function negativity in a classical model of quantum light}

\author{Brian R. {La Cour}}

\address{Applied Research Laboratories, The University of Texas at Austin, \\
P.O. Box 9767, Austin, Texas 78766-9767, United States, \\
blacour@arlut.utexas.edu}

\maketitle


\begin{abstract}
The presence of negative values in the Wigner quasiprobability distribution is deemed one of the hallmarks of nonclassical phenomena in quantum systems.  Here we demonstrate a classical model of squeezed light that, when combined with post-selection on amplitude threshold-crossing detection events, is capable of reproducing observed behavior of single-photon added coherent states.  In particular, a classical model of balanced homodyne detection and standard tomographic techniques are used to infer the density matrix in the Fock basis.  The resulting Wigner functions exhibit negatively for photon-added vacuum and weak coherent states.
\end{abstract}


\section{Introduction}

The Wigner function was first proposed by Eugene Wigner in 1932 as a pseudo probability density function for quantum systems describing the joint distribution of noncommuting observables \cite{Wigner1932}.  It provides a mathematically equivalent description for the quantum state of a physical system and, through the Wigner-Weyl transform, can be put in one-to-one correspondence with the density operator of the state \cite{Weyl1927}.  Although the marginal distributions of quadrature observables are accurately captured by the Wigner function, their joint distribution may require that the Wigner function attain negative values in certain regions of phase space.  As such, the Wigner function cannot always be deemed a proper probability density function.

Many quantum states, such as Gaussian coherent and squeezed states, do have entirely positive Wigner functions and, thus, may be represented as complex random variables.  Others, such as single-photon Fock states, exhibit negativity and, thus, cannot be so represented.  More generally, non-Gaussian pure states have a Wigner function that is necessarily negative, though this constraint is relaxed for mixed states \cite{Hudson1974,Soto1983,Walschaers2021}.  Negativity of the Wigner function is connected to quantum contextuality and has been identified as an important resource for quantum information processing \cite{Raussendorf2017,Lund2008,Andersen2015}.  Such states are therefore of great theoretical as well as practical importance.

A single-photon-added coherent state (SPACS) provides one technique for producing non-classical states of light with Wigner function negativity \cite{Agarwal1991}.  In this method, a beamsplitter is used to mix a coherent light source with a heralded photon from an entangled source pair, typically generated via parametric downconversion \cite{Zavatta2004b,Zavatta2005}.  Balanced homodyne detection with a phase-controlled local oscillator may then be used on the two output ports to measure the resulting quantum state using standard tomographic techniques \cite{Leonhardt,Lvovsky2009}.  Several experiments of this sort have been performed, with the measured Wigner function exhibiting negativity of up to 13\% of the minimum for a single-photon Fock state \cite{Lvovsky2001,Zavatta2004a,Barbieri2010}.

Although a negative Wigner function clearly rules out a possible classical interpretation as a joint distribution of quantum observables, it is nevertheless possible to obtain such a result classically through tomographic inference and post selection.  Prior work has suggested that the following ingredients are needed (1) a model of the vacuum modes as real, not virtual, (2) a nonlinear transformation of the reified modes to represent squeezed states, and (3) a model of photon detection based on amplitude threshold crossing events \cite{LaCour2020}.  These ingredients, combined with the standard techniques for performing and analyzing SPACS experiments, suffice to provide an inferred density matrix whose corresponding Wigner function is capable of exhibiting negativity, as will be shown here.


\section{Classical Model}

\begin{figure}[ht]
\centerline{\scalebox{0.35}{\includegraphics{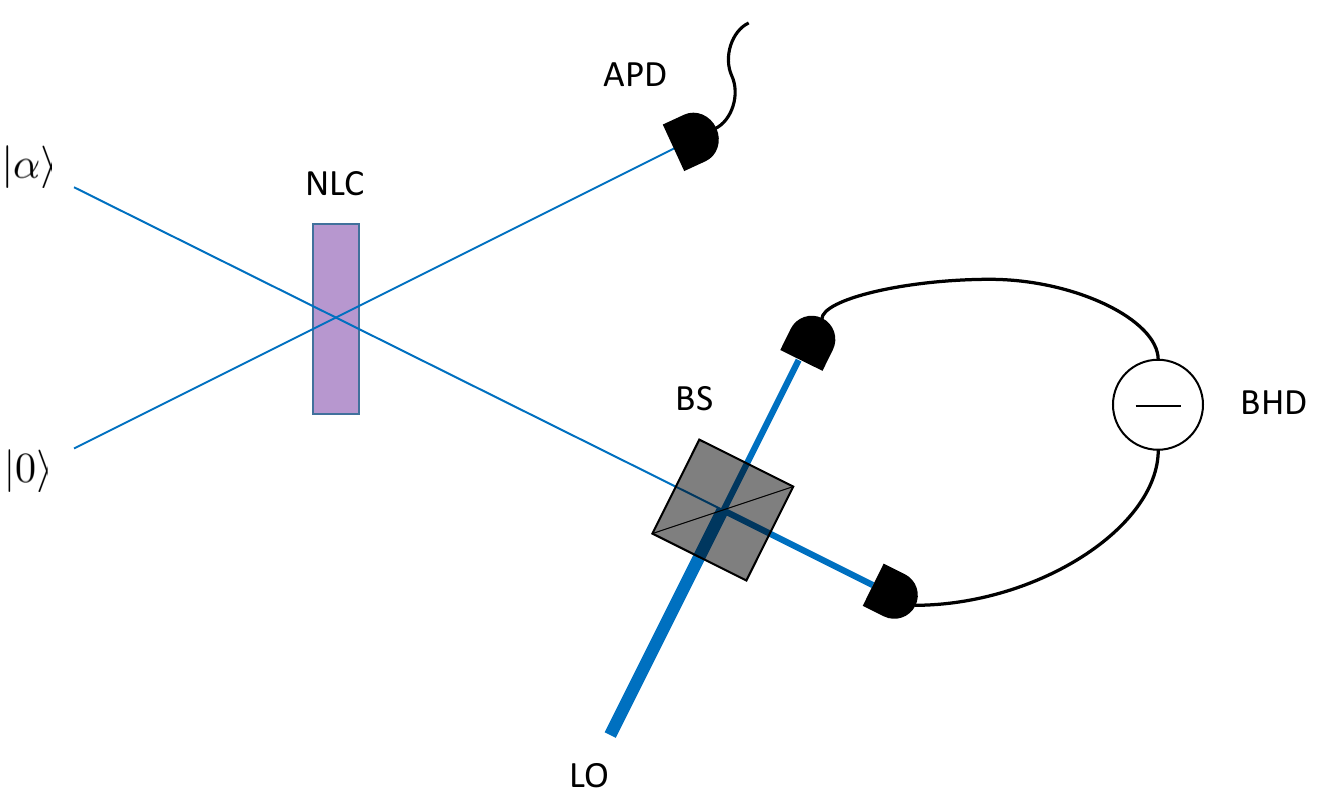}}}
\caption{(color online) Notional experimental configuration for balanced homodyne measurement of a single-photon-added coherent state.  A nonlinear crystal (NLC) is pumped to produce an entangled squeezed state of a coherent state $\ket{\alpha}$ and vacuum mode $\ket{0}$.  An avalanche photondiode (APD) is used to herald the detection of one of the entangled pair, while the other enters a beamsplitter (BS), which is mixed with a phase-controlled local oscillator (LO).  The output ports are then subject to balanced homodyne detection (BHD).}
\label{fig:experiment}
\end{figure}

Figure \ref{fig:experiment} shows the notional setup for a SPACS preparation-and-measurement experiment.  A nonlinear crystal (NLC) is pumped with classical light and mixes with two lower-frequency modes corresponding the coherent signal and vacuum idler states $\ket{\alpha}$ and $\ket{0}$, respectively.  Based on their corresponding Wigner functions, we model these classically as the random variables
\begin{align}
a_s &= \alpha + \sigma z_s \\
a_i &= \sigma z_i \; ,
\end{align}
where $\alpha \in \mathbb{C}$ and $z_s, z_i$ are independent standard complex Gaussian random variables with zero mean and unit variance.  The parameter $\sigma$ describes the scale of the background fluctuations.  For the zero-point vacuum state, $\sigma = 1/\sqrt{2}$, corresponding to a modal energy of $\frac{1}{2} \hbar \omega$, while for thermal states $\sigma > 1/\sqrt{2}$.

The $\chi^{(2)}$ nonlinearity of the crystal transforms the signal and idler states to produce the transformed amplitudes $b_s$ and $b_i$, where 
\begin{align}
b_s &= \cosh(r) \, a_s + \sinh(r) \, a_i^* \\
b_i &= \cosh(r) \, a_i + \sinh(r) \, a_s^*
\end{align}
and $r \ge 0$ is the squeezing parameter corresponding to the strength of the pump and the geometry and nonlinearity of the crystal \cite{Boyd2008}.  These classical equations correspond formally to the Bogoliubov transformations of the corresponding raising and lowering operators for a dual-mode squeezed state \cite{Agarwal}.  We note that $b_s$ and $b_i$ are complex Gaussian random variables, much like $a_s$ and $a_i$; however, for $r > 0$ they will be both correlated and improper, owing to a nonzero covariance and pseudo covariance \cite{Schreier&Scharf}.  The impropriety of their statistical distribution is a reflection of the entangled nature of the two modes \cite{LaCour2021}.

The transformed idler mode is subject to detection via an avalanche photodiode (APD) in order to perform heralding for the signal mode.  This process is modeled classically as a deterministic amplitude threshold crossing event, given by
\begin{equation}
D = \left\{ (z_i, z_s) \in \mathbb{C}^{2} : |b_i| > \gamma \right\} \; ,
\end{equation}
where $\gamma \ge 0$ is a predetermined threshold corresponding to the detector.  Such a model of single-photon detection has been suggested previously and is capable of reproducing many observed quantum phenomena \cite{Adenier2009,Khrennikov,LaCour2014,LaCour2020}.  Let $c_s$ denote the signal amplitude, $b_s$, conditioned on the detection event $D$.

The post-selected signal amplitude, $c_s$, is now subject to balanced homodyne measurements.  To do this, the signal is mixed with a phase-controlled local oscillator (LO) using a 50/50 beamsplitter, and the intensities of the two output ports are measured and subtracted.  If the phase of the local oscillator is $\theta$, this results in a measurement of the quadrature component
\begin{equation}
q_{\theta} = \frac{c_s e^{-i\theta} + c_s^* e^{i\theta}}{\sqrt{2}} \; .
\end{equation}

Unconditioned, $q_{\theta}$ is normally distributed with a mean of $\sqrt{2}\cosh(r) \, \mathrm{Re}[\alpha e^{-i\theta}]$ and a variance of $[2\cosh(r)^2 - 1] \sigma^2$.  We shall, however, be interested in the conditional distribution of this real-valued random variable.  Let $p_{\theta}(q)$ denote the conditional probability density function of $q_{\theta}$, given $D$.  Specifically, let
\begin{equation}
p_{\theta}(q) = \frac{\d}{\d q} \Pr[ \, q_{\theta} \le q \, | \, D \, ] \; .
\end{equation}
This expression will be evaluated numerically in the following section.

Given $p_{\theta}(q)$, the matrix components of the inferred quantum density matrix $\rho$ in the Fock basis are given by
\begin{equation}
\rho_{nm} = \int_{0}^{\pi} \int_{-\infty}^{\infty} p_{\theta}(q) \, f_{nm}(q) \, \d q \, e^{i(n-m)\theta} \, \d \theta \; ,
\label{eqn:rhonm}
\end{equation}
where $f_{nm}(q)$ is the pattern function corresponding to modes $m$ and $n$ \cite{DAriano1994,Leonhardt1996}.  Once these are determined, the Wigner function is easily computed using the expansion
\begin{equation}
W(\alpha) = \sum_{n,m=0}^{\infty} \rho_{nm} \, W_{nm}(\alpha) \; ,
\end{equation}
where $W_{nm}(\alpha)$ is written in terms of the associated Laguerre polynomials as
\begin{equation}
W_{nm}(\alpha) = \frac{2}{\pi} \frac{(-1)^m}{\sqrt{2^{m-n}}} \sqrt{\frac{m!}{n!}} \, \frac{e^{-2|\alpha|^2}}{\alpha^{m-n}} L_m^{n-m}(4|\alpha|^2)
\end{equation}
for $m \le n$ and $W_{nm}(\alpha) = W_{mn}(\alpha)$ otherwise \cite{Curtright}.


\section{Numerical Simulations}

\begin{figure}[ht]
\centerline{\scalebox{0.6}{\includegraphics{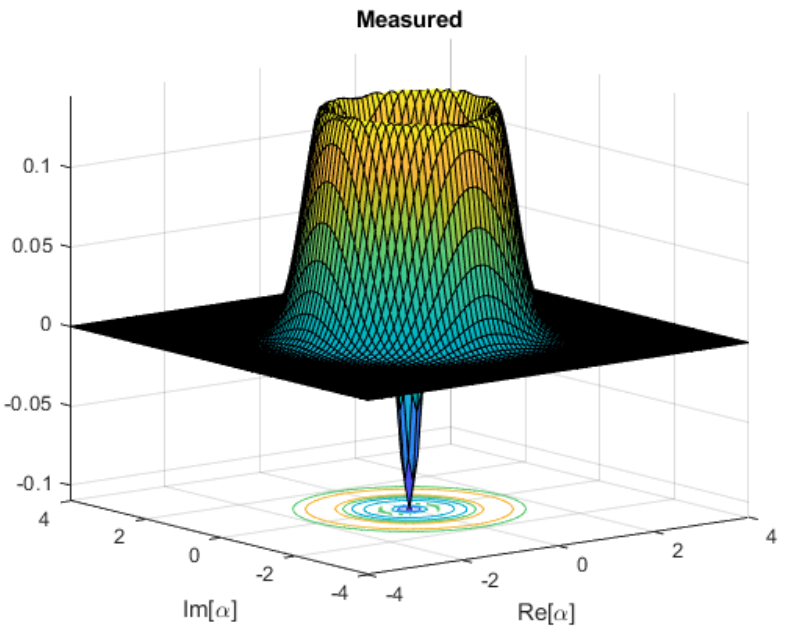}}}
\caption{(color online) Numerically inferred Wigner function for a classically modeled single-photon-added vacuum state.}
\label{fig:Wigner0}
\end{figure}

A series of numerical simulations were performed, taking various values of the parameters $\alpha$, $\sigma$, $\gamma$, and $r$.  For each case, multiple realizations of $z_i$ and $z_s$ were generated such that the conditioning set $D$ contained at least $2^{16}$ elements.  Next, $\theta$ was varied from $0^\circ$ to $180^\circ$ in $1^\circ$ increments, and the corresponding values of $q_{\theta}$ over the conditioning set $D$ were computed.  The resulting histogram provided an estimate of $p_{\theta}(x)$.  These, in turn, were numerically integrated according to Eqn.\ (\ref{eqn:rhonm}) using a numerically stable pattern function routine \cite{Leonhardt1996}.

Figure \ref{fig:Wigner0} shows an example of the inferred Wigner function for the vacuum state ($\alpha = 0$) using $\sigma = 1/\sqrt{2}$, $\gamma = 2.5$, and $r = 0.4$.  A maximum of four modes (i.e., $n,m \in \{0,\ldots,3\}$) were used in the expansion.  The minimum of the function is $-0.11$, which is just over 17\% of the $-2/\pi$ minimum for a single-photon Fock state.

\begin{figure}[ht]
\centerline{\scalebox{0.6}{\includegraphics{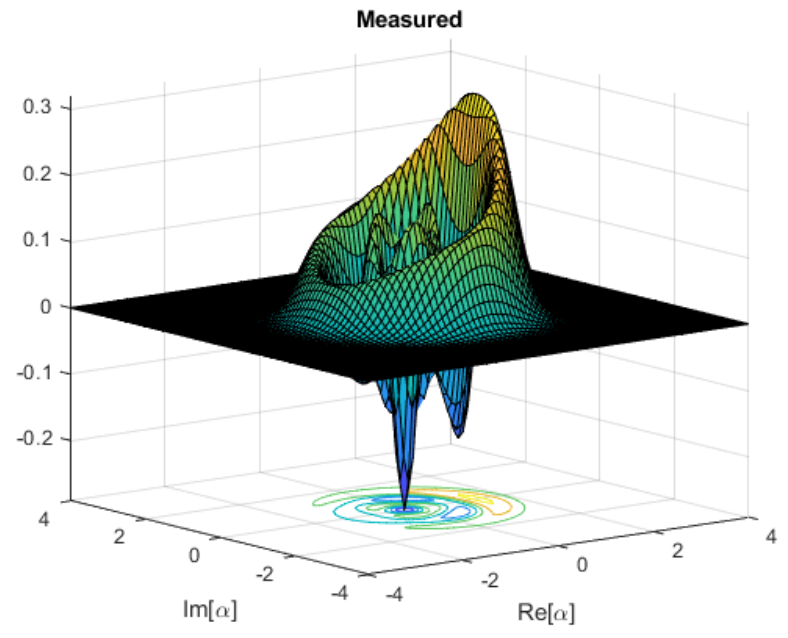}}}
\caption{(color online) Numerically inferred Wigner function for a classically modeled single-photon-added coherent state with an average photon number of one.}
\label{fig:Wigner1}
\end{figure}

Figure \ref{fig:Wigner1} shows a different example using the same parameter settings but with $\alpha = 1$, corresponding to a coherent state with an average photon number of 1.  A peak near $\alpha = 1.4$ is clearly visible, as is the minimum negative value of $-0.29$ at the origin; however, additional local extrema can be seen as well, indicating the presence of higher-order modes.  Relative to the ideal SPAC state, the fidelity is $0.74$.


\section{Conclusions}

With post-selection on heralding events, a wholly classical model of light is capable of producing quadrature measurements whose conditional distributions give rise to inferred Wigner functions that clearly exhibit negativity.  Numerical simulations show good agreement with experimental observations for single-photon-added coherent states but differ from ideal predictions in the presence of higher-order modes.  Such deviations may be attributed to the method of heralding, which does not discriminate the number of photons, as well as the inevitable introduction of higher-order photon number states in the initial multi-mode squeezed state.




\section*{Acknowledgements}

This work was funded by the Office of Naval Research under Grants {N00014-18-1-2107} and {N00014-23-1-2115}.


\section*{ORCID}

\noindent Brian R. La Cour - \url{https://orcid.org/0000-0001-7899-0938}





\end{document}